\documentclass[aps,pre,twocolumn,groupedaddress,showpacs]{revtex4}
\usepackage{graphicx}
\newcommand{\bx}{\mathbf{x}}
\newcommand{\by}{\mathbf{y}}
\newcommand{\bz}{\mathbf{z}}
\newcommand{\bp}{\mathbf{p}}

\newcommand{\be}{\mathbf{e}}

\newcommand{\bg}{\mathbf{g}}

\newcommand{\hx}{\hat{x}}

\newcommand{\hg}{\hat{g}}
\newcommand{\bzero}{{\bf 0}}
\newcommand{\tri}{\triangle}

\newcommand{\bQ}{\mathbf{Q}}

\newcommand{\sep}{ \ \ \ , \ \ \ }

\newcommand{\beq}{\begin{equation}}
\newcommand{\eeq}{\end{equation}}
\newcommand{\beqn}{\begin{eqnarray}}
\newcommand{\eeqn}{\end{eqnarray}}
\newcommand{\pp}{\partial}
\newcommand{\dd}{{\rm d}}
\newcommand{\ee}{{\rm e}}
\newcommand{\eq}{Eq.\ }
\newcommand{\eqs}{Eqs\ }
\newcommand{\fig}{Fig.\ }

\newcommand{\la}{\langle}
\newcommand{\ra}{\rangle}

\newcommand{\app}{Appendix\ }
\newcommand{\cf}{c.f.\ }

\begin{document}

\title{Thermal breakage of a discrete one-dimensional string
}
\author{Chiu Fan \surname{Lee}
}
\email{C.Lee1@physics.ox.ac.uk}
\affiliation{Physics Department, Clarendon Laboratory,
Oxford University, Parks Road, Oxford OX1 3PU, UK}


\date{\today}

\begin{abstract}
We study the thermal breakage of a discrete one-dimensional string, with open and fixed ends, in the heavily damped regime. Basing our analysis 
on the multidimensional Kramers escape theory, we are able to make analytical predictions on the mean breakage rate, and on the breakage propensity with respect to the 
breakage location on the string. We then support our predictions with numerical simulations.
\end{abstract}

\maketitle

\begin{figure}
\caption{(a) A 1D mass-spring system with fixed ends. (b) A 1D mass-spring system with open ends. In both cases, the separation between the beads are assumed to be of 
unit length.
}
\label{pic}
\begin{center}
\includegraphics[scale=.35]{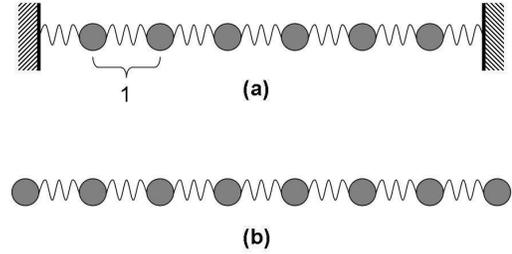}
\end{center}
\end{figure}

Recently, there is much discussion on the possibility of exploiting biopolymers as functional materials 
\cite{Sarikaya_NatMaterials03,Zhang_NatBiotech03,Hartgerink_Science01,Han_AdvMaterials07}. To achieve this goal, the stabilities of such materials have to be thoroughly 
investigated. Furthermore, the facts that the biopolymers are necessarily finite and consist of discrete parts, such as individual peptides in an amyloid fibril 
\cite{Zhang_NatBiotech03}, have to be taken into consideration.
As a step towards this direction, we study here a toy model for the breakage of a discrete one-dimensional string under thermal fluctuations, in both fixed-ended and open-ended configurations (c.f. \fig \ref{pic}). 
This problem has been studied previously by numerical simulations \cite{Welland_PRB92,Oliveira_JChemPhys94,Bolton_JPhysChem95} and theoretically with phenomenological 
assumptions on the effect of friction on the collective modes \cite{Sebastian_ChemPhysLett99,Puthur_PRB02}. Multi-dimensional Kramers escape theory has also been applied 
to the study of breakage in a one-dimensional ring \cite{Sain_PRE06}. 
The energy profile for the bonds in the string is usually modeled by a quadratic potential at the minimum energy region, and by an inverted quadratic potential at the 
breakage point.  Here, we employ a simplified model where all bonds are assumed to be Hookian up to the breakage point. This model has the virtue of rendering the 
theoretical analysis asymptotically exact as temperature goes to zero.
By studying in detail the energy dependency on the collective modes, we are able to employ the multidimensional Kramers escape theory to predict the breakage rate {\it and} the breakage propensity with respect to the breakage location. These predictions are then verified by numerical simulations.

\subsection{String with fixed ends}
We consider the dynamics of a one-dimensional string modeled as a collection of $M$ masses connected by springs with identical spring constant $\kappa$. We also assume 
that we are in the heavily damped regime, i.e., the inertia terms are ignored, which is reasonable for many biopolymers in typical experimental conditions 
\cite{Sneppen_B05}.
We assume that the beads in the string are initially at the minimal energy configuration, i.e., each consecutive pair of beads is separated by a unit distance.  
Denoting the positional deviation of the $n$-th bead from the initial configuration by $x_n$, the equations of motion under thermal perturbation are of the form:
\beq
\label{main}
\frac{\dd \bx}{\dd t} = -\frac{\kappa}{\zeta} A \bx +\bg
\eeq
where $\zeta$ is the damping coefficient and $\bg$ is a Gaussian noise such that
\beqn
\label{g}
\la \bg \ra &=& \bzero
\\
\nonumber
\la g_m(t) g_n(t') \ra &=& \frac{2k_BT}{\zeta} \delta_{mn}\delta(t-t') \ ,
\eeqn
and
\beq
\label{MA1}
A=\left[
\begin{array}{rrrrr}
2 & -1 & 0 & \cdots & 0
\\
-1 & 2 & -1 & \cdots & 0
\\
\vdots & & \ddots & & \vdots
\\
0 & \cdots &0 & -1 & 2
\end{array}
\right]  \ ,
\eeq
i.e., $A_{nn} =2$ and $A_{n+1,n} = A_{n,n+1} = -1$.

As $A$ is symmetric, it is diagonalizable by a set of orthonormal vectors. Let $D$ be the diagonal matrix such that $VDV^\dagger = A$ where $V$ is the corresponding 
orthogonal matrix, we have
\beq
\label{p}
\frac{\dd \bp}{\dd t} = -\frac{\kappa}{\zeta} D \bp +\bg
\eeq
where $\bp = V^\dagger \bx$ and the thermal perturbation term $\bg$ remains the same since $V$ is an orthogonal matrix. The diagonal elements of $D$ are
\beq
\lambda_s = 2 \left[1- \cos \frac{\pi s}{M+1}  \right]
\eeq
and the entries in $V$ are:
\beq
V_{ns} = C\sin \frac{\pi sn}{M+1} \ .
\eeq 
In the above equation, $C = \sqrt{2/(M+1)}$ is the normalization factor so that $\sum_n V_{ns}^2 =1$ for all $s$ (c.f. \app \ref{A1}).

The extension/contraction of the $n$-th spring (designating the spring before the $n$-th bead) is given by:
\beqn
e_1 &=& [V\bp]_1
\\
e_n &=& [V\bp]_n - [V\bp]_{n-1} 
\\
e_{M+1} &=& [V\bp]_M - [V\bp]_{M-1}\ ,
\eeqn
or in matrix notation,
\beq
\be = W \bp
\eeq
where $W$ is a matrix of dimension $(M+1)\times M$ such that
\beqn
\label{W}
W_{1s} &=& V_{1s}
\\
\nonumber
W_{ns} &=& V_{ns} -V_{n-1,s} \sep 1 <n \leq M
\\
\nonumber
W_{M+1,s} &=& V_{M,s}\ .
\eeqn
We are interested in the Mean First Breakage Time (MFBT) defined as:
\beq
\tau = \inf_t \Big\{t > 0\ \Big|\ \max_n \{ |e_n(t)|\} > b \Big\} \ . 
\eeq
In physical terms, we are to find the average waiting time before any of the $e_n$ is  extended or contracted by an amount $b$ where $b<1$ \footnote{Note that the same 
formalism still applies if we consider the case of breakage by extension alone. The resulting MFBT will simply be twice the amount being calculated here.}. The above 
problem is equivalent to a multi-dimensional Kramers escape problem \cite{Langer_AnnPhys69, Hanggi_RMP90}, and we will employ the formalism developed in \cite{Matkowsky_SIAMJApplMath77} for our analysis. 

In terms of the normal modes $\{ p \}$, the total energy corresponding to the entire string is 
\beq
U(\bp) = \frac{ \kappa}{2}\sum_s \lambda_s p_s^2 \ ,
\eeq
and the exit boundaries are defined by:
\beq
\sum_s W_{ns} p_s =\pm b \sep {\rm for \ } 1 \leq n \leq M+1 \ .
\eeq
Our first task  is to find the exit routes with the minimal energy. To do so, we  employ the Lagrange multiplier method to minimize the following quantity with respect to 
$\bp$:
\beq
U(\bp) + z_n \left(\sum_s W_{ns} p_s \pm b \right) \sep {\rm for \ } 1 \leq n \leq M+1
\eeq
where $z_n$ is the corresponding Lagrange multiplier.
The solution to this minimization problem is that for each $n$, there are two minimizing vectors, $\hat{\bf p}^{(n)+}$ and $\hat{\bf p}^{(n)-}$, of the forms:
\beq
\hat{p}_s^{(n)\pm} = \pm \frac{W_{ns}b}{\lambda_s}\left( \sum_s \frac{W_{ns}^2}{\lambda_s} \right)^{-1} \ , 
\eeq
such that the corresponding energy is:
\beq
U(\hat{\bf p}^{(n)\pm}) = \frac{\kappa b^2}{2}\left( \sum_s \frac{W_{ns}^2}{\lambda_s} \right)^{-1} \ .
\eeq

\begin{figure}
\caption{(Color online) Consider the case of having three beads connected by one black and one red spring. The energy profile in the $p$-space is depicted in the surface plot in the lower-left figure. Breakages of the black (red) spring correspond to the black (red) boundaries on the energy plot. A possible trajectory that leads to breaking the black spring is schematically depicted in white. The energy profile at the upper-right exit boundary is depicted in the lower-right figure. The MFBT is partly determined by the curvature of the potential energy at the exit point (\cf \eq (\ref{tau})).
}
\label{pic2}
\begin{center}
\includegraphics[scale=.4]{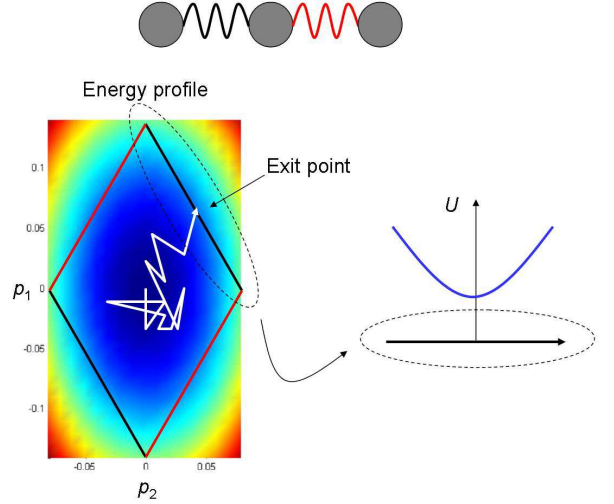}
\end{center}
\end{figure}

Since (c.f. \app \ref{A2})
\beq
\sum_s \frac{W_{ns}^2}{\lambda_s}  = \frac{M}{M+1} \sep 1\leq n \leq M+1 \ ,
\eeq
the minimal energy is the same for all $n$:
\beq
 U(\hat{\bf p}^{(n)\pm}) = \frac{\kappa b^2 (M+1)}{2  M}  \sep 1\leq n \leq M+1 \ .
\eeq
As a result, there are $2M+2$ exit points at the boundary that correspond to the same energy. We will denote these exit points by $\bQ^{(k)}$:
\beq
Q_s^{(k)}= (\pm 1)^k \frac{W_{\lceil k/2 \rceil ,s } b (M+1)}{\lambda_s M} \sep 1\leq k \leq 2M+2 \ .
\eeq
In \fig \ref{pic2}, we depict the energy profile and the corresponding exit points for the case of a three-bead system.

With the formalism developed in \cite{Matkowsky_SIAMJApplMath77}, in the asymptotic limit of $\kappa \rightarrow \infty$ (or equivalently, $k_BT \rightarrow 0$),
the MFBT can be expressed exactly as:
\beq
\label{tau}
\tau=\frac{\zeta \sqrt{2 \pi k_BT} \ \exp \left[\frac{(M+1)\kappa b^2}{2M k_BT} \right]}{ \phi_0^{1/2} \sum_{k=1}^{2M+2}\phi_{\bar{k}}^{-1/2} |\kappa D\bQ^{(k)}|}
\eeq
where
\beqn
\label{phi0}
\phi_0 &=& \det \left. \frac{\pp^2 U }{\pp p_r \pp p_s} \right|_{\bp = \bzero}
= \kappa^M \prod_s \lambda_s
\\
\phi_{\bar{k}} &=&  \det \left. \frac{\pp^2 U }{\pp \hat{p}_r \pp \hat{p}_s} \right|_{\bp = \bQ^{(k)}} \sep 1\leq r,s \leq M-1
 \ ,
\eeqn
with $\hat{p}_r (1 \leq r \leq M-1)$ being a set of basis that are perpendicular to the direction of the $k$-th exit route, i.e., they are perpendicular to the direction $(W_{\lceil k/2 \rceil ,1} , \ldots, W_{\lceil k/2 \rceil ,M} )$.
In other words, $\phi_0$ corresponds to the Hessian of the potential energy at the origin, and $\phi_{\bar{k}}$ corresponds to the Hessian of the potential energy within the hyperplane that has its normal pointing along the $k$-th exit route. Note also that physically,
$ |\kappa D\bQ^{(k)}|$ corresponds to the magnitude of the potential energy's gradient at the $k$-th exit point
\cite{Matkowsky_SIAMJApplMath77}.

\begin{figure}
\caption{(Color online) Breakage events for the fixed-ended string. {\it Upper plot:} The ratios of the MFBT from simulations {\it vs}.\ the MFBT from theory. Each 
marker represents 1000 samples. The parameters are: $\zeta =10$, $k_BT =1$ and $b=0.1$. {\it Lower plot:} The breakage frequency with respect to the breakage location for the case of $M=8$ with the combined data from the cases of $\kappa = 1600, 1800, 2000$.
}
\label{ResA}
\begin{center}
\includegraphics[scale=.6]{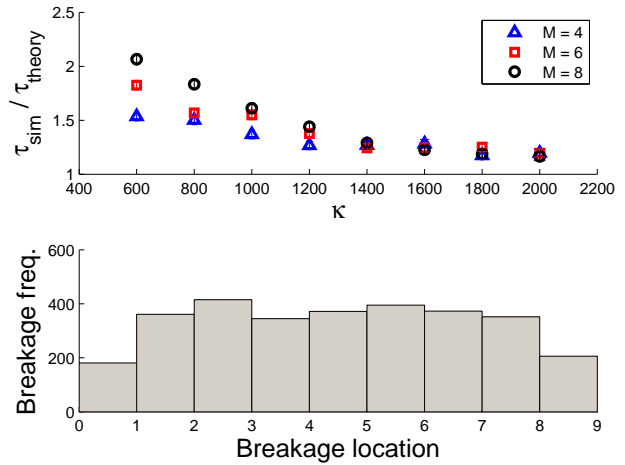}
\end{center}
\end{figure}

In \app \ref{A3} (\eq (\ref{Q})) and \app \ref{hessian} (for \eqs (\ref{p0}) and (\ref{pk})), we argue that 
\beqn
\label{Q}
\left|\frac{\kappa}{\zeta} D\bQ^{(k)} \right| &=& 
\left\{
\begin{array}{l}
 \frac{(M+1)\kappa b}{\zeta  M}\sep  k=1,2,2M+1,2M+2
\\
\frac{\sqrt{2}(M+1)\kappa b}{\zeta  M}\sep  {\rm otherwise}
\end{array}
\right.
\\
\label{p0}
\phi_0  &= & \kappa^M(M+1)
\\
\label{pk}
\phi_{\bar{k}}&=& 
\left\{
\begin{array}{l}
\kappa^{M-1}M \sep  k=1,2,2M+1,2M+2
\\
\kappa^{M-1}M/2 \sep  {\rm otherwise} \ .
\end{array}
\right.
\eeqn
Substituting the above quantities to \eq (\ref{tau}), we arrive at the following prediction for the MFBT:
\beq
\tau = \sqrt {\frac{ \pi k_BTM}{8(M+1)}} \frac{ \zeta}{\kappa^{3/2} b (M+1) }  \ \exp\left[ \frac{(M+1)\kappa b^2}{2k_BTM} \right]\ .
\eeq
Fig.~\ref{ResA} demonstrated that the convergence of the numerical results to the analytical predictions as $\kappa$ increases. 

Besides the MFBT, this formalism is also capable of predicting the propensity for breakage with respect to the location of the spring in the string. According to \eq 5.1 
in \cite{Matkowsky_SIAMJApplMath77}, the probability of breakage at the $n$-th segment is given by:
\beqn
\label{Pr}
\Pr(n) &=&  \frac{2\phi_{\bar{k}}^{-1/2}|\kappa D\bQ^{(n)}/\zeta|}{\sum_{k=1}^{2M+2}  \phi_{\bar{k}}^{-1/2}|\kappa D\bQ^{(k)}/\zeta|}
\\
&=& \left\{
\begin{array}{ll}
1/(2M) \sep & n=1,M+1
\\
1/M  \sep & {\rm otherwise} \ .
\end{array}
\right.
\eeqn
This signifies that the breakage propensity is uniform for all springs except for the two
extremal springs, which break half as frequently as the springs in the middle. \fig \ref{ResA} shows that the analytical predictions are in good 
agreement with simulations. Physically, the fact that the extremal springs break half as  frequently may be seen from the fact that they are connected to the rigid wall 
on one side and so they are subjected to about half the amount of thermal fluctuations.

\begin{figure}
\caption{(Color online) Breakage events for the opened-ended string. {\it Upper plot:} The ratios of the MFBT from simulations {\it vs}.\ the MFBT from theory. Each %
marker represents 1000 samples. The parameters are: $\zeta =10$, $k_BT =1$ and $b=0.1$. {\it Lower plot:} The breakage frequency with respect to the breakage location for the case of $M=8$ with the combined data from the cases of $\kappa = 2000, 2200, 2400$.
}
\label{ResB}
\begin{center}
\includegraphics[scale=.6]{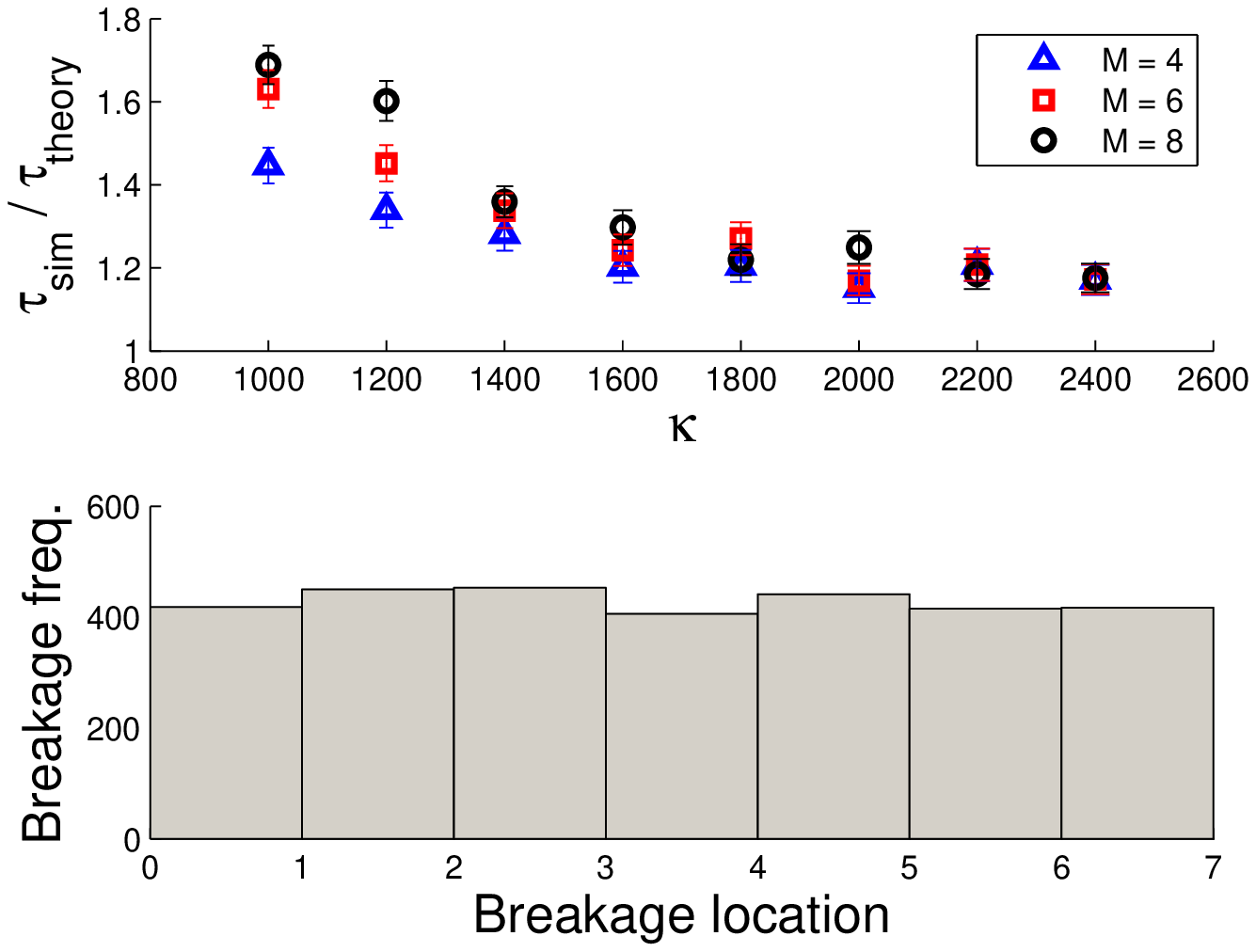}
\end{center}
\end{figure}

\subsection{Strings with open ends}
For an open-ended string as depicted in \fig \ref{pic}b, the equations of motions are governed by:
\beqn
\label{B1}
\frac{\dd \hx_1}{\dd t} &=&
\hx_2 -\hx_1 -1 + \hg_1
\\
\nonumber
\frac{\dd \hx_n}{\dd t} &=&
\hx_{n+1} -2\hx_n +\hx_{n-1}  +\hg_n
\sep 1 < n< M
\\
\nonumber
\frac{\dd \hx_M}{\dd t} &=&
\hx_{M-1} -\hx_M +1 +\hg_M
\eeqn
where $\hx_n(t=0) = n$ and  $\bg$ is as defined in \eq (\ref{g}). We now let 
\beq
x_n =\hx_n + \frac{M-1}{2} -n+1 -\frac{1}{M} \sum_n \hx_n  \ ,
\eeq
i.e., these new coordinates are defined in relation to the center of mass of the string.
With these transformations, the equations of motion become
\beq
\frac{\dd \bx}{\dd t} = -\frac{\kappa}{ \zeta}A\bx +\bg
\eeq 
where
\beq
\label{MA2}
A=\left[
\begin{array}{rrrrr}
1 & -1 & 0 & \cdots & 0
\\
-1 & 2 & -1 & \cdots & 0
\\
\vdots & & \ddots & & \vdots
\\
0 & \cdots &0 & -1 & 1
\end{array}
\right] 
\eeq
with $\bg$ remains the same as in \eq (\ref{g}).

With $VDV^\dagger =A$, the diagonal elements of D are 
\beq
\lambda_s = 2 \left[1- \cos \frac{\pi (s-1)}{M}  \right]
\eeq
and $V$ is now defined by
\beq
V_{ns} = C\cos \frac{\pi (n-1/2)(s-1)}{M} \ .
\eeq 
In the above equation,  $C = \sqrt{2/(M+1)}$ is again the normalization factor as in the case of fixed-ended string. The proof for this statement is similar to that 
presented in \app \ref{A1} and is thus 
omitted.

Note that the first normal mode, $p_1$, corresponds to all beads moving in unison and is thus of no interest in terms of breakage events. We will therefore omit this mode in subsequent discussion. In other words, we will only consider the set of normal modes $\{ p_s:2\leq s \leq M\}$.

As in the previous section, we are interested in the extension/contraction given by:
\beq
\be = W \bp
\eeq
where $W$ is a matrix of dimension $(M-1)\times (M-1)$ such that
\beq
W_{ns} = V_{s,n+1} -V_{s,n} \sep 1 \leq n < M ,\ 2\leq s \leq M \ .
\eeq
Here, we have again
\beq
\sum_{s=2}^M \frac{W_{ns}^2}{\lambda_s}  = 1 \sep 1\leq n \leq M-1 \ .
\eeq

For an open-ended string, \eqs (\ref{Q}), (\ref{p0}) and (\ref{pk}) in the previous section are modified to:
\beqn
\left|\kappa D\bQ^{(k)} \right| &=& \sqrt{2} \kappa b
\\
\phi_0 &=& \kappa^{M-1}M
\\
\phi_{\bar{k}}&=& \kappa^{M-2}
\frac{M}{2 }
\ .
\eeqn
The demonstrations of the above equalities are very similar to those presented in \app \ref{A3}, \ref{A4} and \ref{hessian} are therefore omitted.

Employing \eq (\ref{tau}), we arrive at the following asymptotic (as $\kappa \rightarrow \infty$) prediction for the MFBT 
\beq
\tau = \sqrt {\frac{ \pi k_BT}{8}} \frac{ \zeta}{\kappa^{3/2} b (M-1) }  \ \ee^{\kappa b^2/2k_BT}\ .
\eeq
This prediction is in good agreement with simulations as shown in \fig \ref{ResB}. Note also that according to this calculation, the breakage rate of an open-ended 
string is the same as that of a fixed-ended string when $M \gg 1$.

Since $\phi_{\bar{k}}$ and $\left|\kappa D\bQ^{(k)} \right|$ are identical for all $k$, \eq (\ref{Pr}) predicts that all the springs are broken with equal frequency. In other words, contrary to the case of an 
fixed-ended string, we would expect a flat distribution of breakage frequencies across the string, which is indeed shown to be the case by simulations (c.f. 
\fig~\ref{ResB}). This result may be expected as, unlike in the case of a string with fixed ends, the extremal springs here are again connecting two fluctuating beads. We 
therefore  expect that their breakage frequencies would be similar to those for the springs in the middle of the chain.

\vspace{.2in}
In summary, we have investigated analytically the breakage rate for an 1D string under thermal fluctuation in the heavily damped regime. Our 
approach is based on the theory of the multi-dimensional Kramers escape problem, and we have supported our analytical predictions with  numerical simulations.

\begin{acknowledgements}
The author thanks the Glasstone Trust (Oxford) and Jesus College (Oxford) for financial support. 
\end{acknowledgements}

\appendix
\section{Proofs of various identities}
\label{proofs}

\subsection{Claim: $C^2  = 2/(M+1)$}
\label{A1}
By definition:
\beqn
C^{-2} &=& \sum_s \sin^2\frac{\pi sn}{M+1} 
\\
&=& \frac{1}{2} \sum_{s=1}^M \left(1  - \cos \frac{2\pi ns}{M+1} \right)
\\
\label{cos}
&=& \frac{M+1}{2} - \frac{1}{2}\sum_{s=0}^M \cos \frac{2\pi ns}{M+1}
\eeqn
where the second term in \eq (\ref{cos}) is zero as the negative terms cancel the positive terms exactly.

\subsection{Claim: $  \sum_s \frac{W_{ns}^2}{\lambda_s}=1$ for all $n$}
\label{A2}

For $n=1$, we have 
\beqn
 \sum_s \frac{W_{ns}^2}{\lambda_s}&=& \frac{C^2}{2}\sum_{s=1}^M \frac{\sin^2(\pi s/(M+1))}{[1-\cos(\pi s/(M+1)]}
\\
&=& \frac{C^2}{2}\sum_s  [1+\cos(\pi s/(M+1)]
\\
&=&\frac{M}{M+1}  \ .
\eeqn
 The case for $n=M$ follows similarly.

If $n$ is not 1 or $M$, $\sum_s \frac{W_{ns}^2}{\lambda_s}$ is by definition:
\beqn
\nonumber
 && 	\frac{C^2}{2}\sum_{s=1}^M \frac{[\sin (\pi sn/(M+1))
 -\sin( \pi s (n-1)/(M+1))]^2
}{1-\cos(\pi s/(M+1)}
\\
\nonumber
&=&
C^2 \sum_s \frac{ \cos^2(\pi s(2n-1)/(2M+2))\sin^2(\pi s/(2M+2))}{ \sin^2(\pi s/(2M+2)}
\\
\nonumber
&=& 
C^2 \sum_s  \cos^2 \frac{\pi s(2n-1)}{2M+2}
\\
\nonumber
&=&
\frac{M}{M+1} + \frac{C^2}{2}\sum_s \cos \frac{\pi s (2n-1)}{M+1} 
\eeqn
as the second term in the last equality is zero.

\subsection{Claim: $\sum_s W^2_{ns}$ equals 1 for $n=1,M+1$ and equals 2 otherwise}
\label{A3}
Firstly we prove that $\sum_s W^2_{ns}$ is 1 for $n=1,M+1$ and is 2 otherwise. The fact that  $\sum_s W^2_{ns}=1$ for $n=1,M+1$ follows immediately from \eqs \ref{W}. 
For $n \neq 1, M+1$, We have
\beqn
\sum_s W^2_{ns} &=&  C^2\sum_s \left[\sin\frac{\pi s n}{M+1}- 
\sin\frac{\pi s (n-1)}{M+1}\right]^2
\\
\nonumber
&=& 2-2C^2 \sum_s \sin\frac{\pi s n}{M+1}
\sin\frac{\pi s (n-1)}{M+1}
\\
\nonumber
&=&2+C^2 \sum_s \left[ \cos \frac{\pi s (2n-1)}{M+1} -\cos \frac{\pi s}{M+1} \right]
\\
\nonumber
&=&2\ .
\eeqn

\subsection{Computations for $\phi_0$}
\label{A4}

By \eq (\ref{phi0}),
\beqn
\phi_0 &=& \kappa^M \prod_{s=1}^M \lambda_s 
\\
&=& (2\kappa)^M \prod_{s=1}^M \left[1 - \cos \frac{\pi s}{M+1} \right]
\\
&=& (4\kappa)^M \left[\prod_{s=1}^M \sin \frac{\pi s}{2(M+1)} \right]^2
\\
&=& \kappa^M(M+1)
\eeqn
where the last equality follows from the identity in \cite{wolfram}.

\subsection{Computations for $\phi_{\bar{k}}$}

\label{hessian}

By rotating the basis, we have a new coordinates, $\by = (y_1, \ldots, y_{M})$, such that the direction $(0, \ldots, 0, 1)$ corresponds to the direction that is normal to the exit boundary at the $k$-th exit point. Let us denote the orthogonal matrix that transforms from this new set of basis back to the $p$-space by $R$. In particular, we have 
\beq
 R_{sM} = \frac{W_{ks} }{\sqrt{\sum_s W_{ks}^2}}\ .
 \eeq
Note that one way to obtain the rest of the matrix elements in $R$ is to employ the Gram-Schmidt process. 
 
In terms of this new coordinates, the potential energy is:
\beqn
U(\by) &=& \frac{ \kappa}{2} \bp^T D \bp
\\
&=&
\frac{ \kappa}{2} \by^T R^TD R \by
\eeqn
Restricting to the first $M-1$ dimensions, $\phi_{\bar{k}}$ is defined as:
\beq
\phi_{\bar{k}} = \det \Big(   [R^TDR]_{1\leq r,s \leq M-1}      \Big)
\ .
\eeq
We have calculated this quantity for $M=3,4, \ldots, 50$ numerically and we find that the following formula is exactly satisfied (up to machine rounding errors):
\beq
\phi_{\bar{k}} = \left\{
\begin{array}{l}
\kappa^{M-1}M \sep  k=1,2,2M+1,2M+2
\\
\kappa^{M-1}M/2 \sep  {\rm otherwise}
\end{array}
\right.
\ .
\eeq
Although we are not able to prove the above equality mathematically, we believe that it holds true for all $M$.

\section{Details of numerical simulations}
\label{simulations}
Simulations performed for the dynamics of strings with fixed ends (free ends) are based on \eq (\ref{main}) with the matrix $A$ defined in \eq (\ref{MA1}) (\eq 
(\ref{MA2})). Namely, the positions of the beads are updated according to the following scheme:
\beq
 \bx (t+\tri t)= \bx(t) -\frac{\kappa}{\zeta}A \bx(t) \tri t + \sqrt{\frac{2k_BT \tri t}{\zeta}}\
 \bz(t)
 \eeq
 where $\bz(t)$ is a vector with entries given by random numbers drawn from the normal distribution with zero mean and a standard deviation of one. 
The simulations always start at the minimal energy configurations and are terminated when one of the springs' lengths become more than 1.1 or smaller than 0.9. 
The parameters in the simulations are: $\zeta=10$, $k_BT=1$, and $\tri t$ is set to be $2\times 10^{-6}$. For each set 
of parameters, 1000 runs are performed.


\end{document}